\documentclass[conference]{IEEEtran}

\IEEEoverridecommandlockouts
\usepackage{cite}
\usepackage{threeparttable}
\usepackage{setspace}
\usepackage{float}
\usepackage{amsmath,amssymb,amsfonts}
\usepackage[inkscapelatex=false]{svg}
\usepackage[colorlinks,urlcolor=blue]{hyperref}
\usepackage{blindtext}
\usepackage{multirow}
\usepackage{algorithmic}
\usepackage{graphicx}
\usepackage{textcomp}
\usepackage{xcolor}
\def\BibTeX{{\rm B\kern-.05em{\sc i\kern-.025em b}\kern-.08em
    T\kern-.1667em\lower.7ex\hbox{E}\kern-.125emX}}

\title{TANGO: A Robust Qubit Mapping Algorithm via Two-Stage Search and Bidirectional Look
}
\DeclareRobustCommand*{\IEEEauthorrefmark}[1]{%
    \raisebox{0pt}[0pt][0pt]{\textsuperscript{\footnotesize\ensuremath{#1}}}}

\author{\IEEEauthorblockN{Kang Xu\IEEEauthorrefmark{1,2},
Yukun Wang\IEEEauthorrefmark{1,2,*}, Dandan Li\IEEEauthorrefmark{3} }
\IEEEauthorblockA{\IEEEauthorrefmark{1}Beijing Key Laboratory of Petroleum Data Mining, China University of Petroleum, Beijing 102249, China \\
\IEEEauthorrefmark{2}Key Lab of Processors, Institute of Computing Technology, CAS, Beijing 100190, China\\
\IEEEauthorrefmark{3}School of Computer Science (National
Pilot Software Engineering School), 
Beijing \\University of Posts and
Telecommunications, Beijing 100876,
China\\
\IEEEauthorrefmark{*}\href{mailto:wykun06@gmail.com}{wykun06@gmail.com}}
}

\begin{document}
\maketitle

\begin{abstract}
Current quantum devices typically lack full qubit connectivity, making it difficult to directly execute logical circuits on quantum devices. This limitation necessitates quantum circuit mapping algorithms to insert SWAP gates, dynamically remapping logical qubits to physical qubits and transforming logical circuits into physical circuits that comply with device connectivity constraints. However, the insertion of SWAP gates increases both the gate count and circuit depth, ultimately reducing the fidelity of quantum algorithms. To achieve a balanced optimization of these two objectives, we propose the TANGO algorithm. By incorporating a layer-weight allocation strategy, the algorithm first formulates an evaluation function that balances the impact of qubit mapping on both mapped and unmapped nodes, thereby enhancing the quality of the initial mapping. Next, we design an innovative two-stage routing algorithm that prioritizes the number of executable gates as the primary evaluation metric while also considering quantum gate distance, circuit depth, and a novel bidirectional-look SWAP strategy, which optimizes SWAP gate selection in conjunction with preceding gates, improving the effectiveness of the mapping algorithm. Finally, by integrating advanced quantum gate optimization techniques, the algorithm's overall performance is further enhanced. Experimental results demonstrate that, compared to state-of-the-art methods, the proposed algorithm achieves multi-objective co-optimization of gate count and circuit depth across various benchmarks and quantum devices, exhibiting significant performance advantages.

\end{abstract}


\section{Introduction}
Quantum computing is one of the most significant approaches in the post-Moore era. It has been widely applied to problems such as integer factorization, database searching, and solving linear systems of equations, demonstrating computational advantages over classical computers in specific tasks and achieving remarkable speedups. Currently, mainstream quantum computing solutions are provided by companies such as IBM\cite{IBM1272021ibm}, Google\cite{google542019quantum}, and Intel\cite{Intel2018ces}, featuring qubit counts of 127, 54, and 49, respectively. In 2024, Google announced the launch of its high-performance quantum processing unit (QPU), Willow\cite{acharya2024quantum}. Compared to its predecessors, this processor achieves significantly reduced error rates, further paving the way for practical quantum computing.

Currently, the dominant model in quantum computing is the quantum circuit model, where each quantum operation is represented as a quantum gate. A typical quantum circuit consists of single-qubit gates and two-qubit gates, with the latter being executable only between directly connected qubits. In Noisy Intermediate-Scale Quantum (NISQ) devices, the connectivity between qubits is often restricted. In order to enable a given quantum circuit to execute on a connectivity-constrained quantum architecture, the original ideal quantum circuit must be transformed. This is typically achieved by inserting SWAP gates, which swap qubits to make non-adjacent qubits physically adjacent, allowing the corresponding quantum gates to be performed. This process, referred to as the qubit mapping problem, primarily aims to minimize the number of SWAP gates introduced during the mapping. This is crucial because the insertion of SWAP gates increases both the gate count and circuit depth, leading to longer runtime and additional noise, which ultimately diminishes the success rate of quantum algorithm execution.
%The complexity of the qubit mapping problem primarily arises from the connectivity limitations of quantum computer architectures and their relationship to quantum gates.%Under ideal conditions, gate operations can be performed directly between all qubits, but in practical quantum computers, due to connectivity constraints, some quantum gates must be performed indirectly by inserting additional SWAP gates. 
%These mapping algorithms typically employ various optimization strategies to balance circuit depth and gate operation counts, aiming to improve the execution efficiency and accuracy of the transformed circuit.

To address this issue, numerous effective qubit mapping algorithms have emerged over the past few decades. Zulehner et al. \cite{zulehner2018efficient} divided the circuit into layers and employed an A* search to find optimal SWAP gates, though its time complexity grows exponentially with search depth. To mitigate this, Li et al. \cite{li2019tackling} introduced SABRE, a bidirectional traversal algorithm with a decay-based heuristic to reduce circuit depth and time complexity. Niu et al. \cite{niu2020hardware} enhanced SABRE with simulated annealing to optimize initial mapping while considering gate noise, which effectively reduced additional CNOT gates and improved circuit fidelity. Furthermore, Zhang et al. \cite{zhang2021time} augmented the A* algorithm with node expansion and pruning operations, and designed the depth-focused heuristics to minimize circuit depth. Chang et al. \cite{chang2021mapping} proposed a novel hexagonal architecture-based mapping algorithm, but its weight assignment only considers individual gate counts, neglecting prioritization between mapped and unmapped nodes, potentially affecting initial mapping quality.

Besides, Liu et al. \cite{liu2022not} proposed NASSC, an optimization-aware routing method that reduces redundant gates via logic synthesis and gate cancellation, outperforming SABRE in both circuit depth and gate count. On the other hand, Li et al. \cite{sqgm2023single} proposed SQGM to optimize the insertion of SWAP gates by storing single-qubit gates, achieving better circuit depth than NASSC when integrated with SABRE but lagging in CNOT gate count. McKinney et al. \cite{mckinney2024mirage} introduced the MIRAGE framework, which reduces SWAP gates by leveraging mirror gates and optimizing their decomposition into the iSWAP family, offering new insights for future quantum circuit mapping. Tan et al. \cite{tan2024compilation} proposed a hybrid strategy for Dynamically Field-Programmable Qubit Arrays (DPQA), combining greedy and optimal methods based on gate count. Their work shows that DPQA-based circuits have lower expansion overheads than fixed-grid architectures, advancing programmable quantum circuits for neutral atom quantum computers. In conclusion, The primary optimization goals of these algorithms typically include minimizing the number of inserted SWAP gates\cite{CNOTwille2016look,SAzhou2020quantum,Fidls2020qubit,ye2024mutual,liu2023tackling,numjiang2024qubit,qian2023method,numpark2022fast,numzhu2021iterated,huang2024efficientSWAP}, reducing quantum circuit depth\cite{depthlao2021timing,depthli2023timing,escofet2024route,fu2023effective}, and lowering the overall error rate of the circuit\cite{noisemurali2019noise,noisetannu2019not,noisedeng2020codar,noisezhu2023variation,huang2024ctqr,noiseash2019qure}. However, existing methods rarely balance the collaborative optimization of quantum gate count and circuit depth, and they lack the capability for further re-optimization of quantum circuits, thus presenting a significant direction for future research.

In this paper, we propose TANGO, a method for jointly optimizing quantum circuit gate count and depth. First, a discriminative function is formulated, incorporating a parameter that balances the impact of qubit placement on both mapped and unmapped nodes, thereby optimizing the initial mapping selection. Then, we propose an innovative two-stage routing algorithm, with the number of executable gates as the primary evaluation metric. Quantum gate distance, circuit depth, and a novel bidirectional look SWAP strategy are subsequently considered as auxiliary evaluation metrics, effectively identifying SWAP gates that achieve optimal exchange efficiency and minimal circuit depth, while also enhancing the potential for subsequent gate cancellation optimization. Finally, by integrating advanced quantum gate optimization techniques, the total gate count and circuit depth are further reduced. Experimental results demonstrate that the algorithm achieves multi-objective co-optimization of gate count and circuit depth across multiple large-scale datasets, further confirming its efficiency and broad applicability. 

The structure of this paper is as follows: Sec. \ref{sec2} introduces the basic concepts and background of the circuit mapping problem; In Sec. \ref{sec3}, the co-optimization mapping algorithms are presented, including the proposed initial mapping method and the innovative two-stage routing algorithm; Sec. \ref{sec4} evaluates the algorithm's performance and analyzes the experimental results; The key findings and contributions of this work are concluded in Sec. \ref{sec5}.

\section{Preliminaries}\label{sec2}

\subsection{Quantum Computing and Quantum Gate}
Quantum computing utilizes qubits, which differ from classical bits that are either 0 or 1. A qubit can be in states $|0\rangle$ or $|1\rangle$, or a superposition of both, expressed as $|\phi\rangle = \alpha|0\rangle + \beta|1\rangle$. Here, $\alpha$ and $\beta$ are complex numbers, and they satisfy the normalization condition $|\alpha|^2 + |\beta|^2 = 1$. 

\begin{figure}[h]
  \centering
  \includegraphics{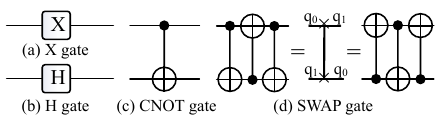}
  \caption{Basic quantum gates in quantum computing}
  \label{basic_gates}
\end{figure}

Quantum gates are the fundamental operations used to manipulate quantum states. Single-qubit gates act on an individual qubit. For example, as shown in Fig. \ref{basic_gates}, the $X$ gate flips the states: $X|0\rangle = |1\rangle$ and $X|1\rangle = |0\rangle$. The Hadamard($H$) gate generates states of equal superposition: $H|0\rangle = (|0\rangle + |1\rangle)/\sqrt{2}$ and $H|1\rangle = (|0\rangle - |1\rangle)/\sqrt{2}$. Two-qubit gates, on the other hand, act on a pair of qubits simultaneously. The most common example is the CNOT (CX) gate, which applies an $X$ gate to the target qubit if the control is $|1\rangle$ state. The SWAP gate exchanges the states of two qubits and can be decomposed into three CNOT gates. In addition to these, there are various higher-dimensional quantum gates, which can ultimately be decomposed into a linear combination of single-qubit and two-qubit gates.

\subsection{Quantum Device and Quantum Circuit}
\begin{figure}[h]
  \centering
  \includegraphics[width=\columnwidth]{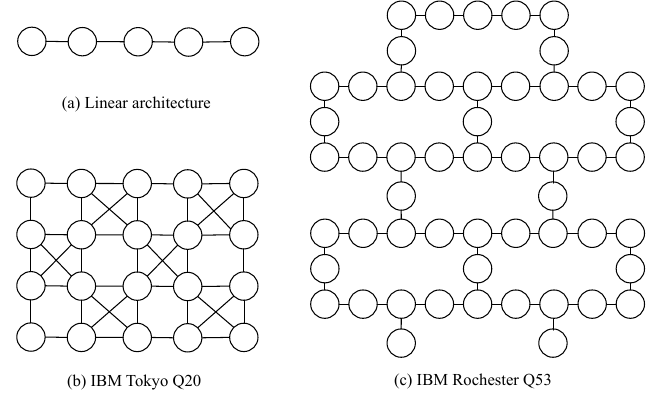}
  \caption{The coupling graphs of real quantum devices}
  \label{architecture}
\end{figure}

The quantum device is represented as \( AG(Q_p, E) \), where \( Q_p \) are physical qubits, and \( E \) are connections between them. In NISQ-era quantum architectures (see Fig. \ref{architecture}), each node represents a physical qubit. It is evident that two-qubit gates can only be performed between physically connected qubits, while single-qubit gates don't. Among these architectures, the IBM  Q20 architecture, with its dense connectivity, is often selected to validate the performance of mapping algorithms.

\begin{figure*}[ht]
  \centering
  \includegraphics{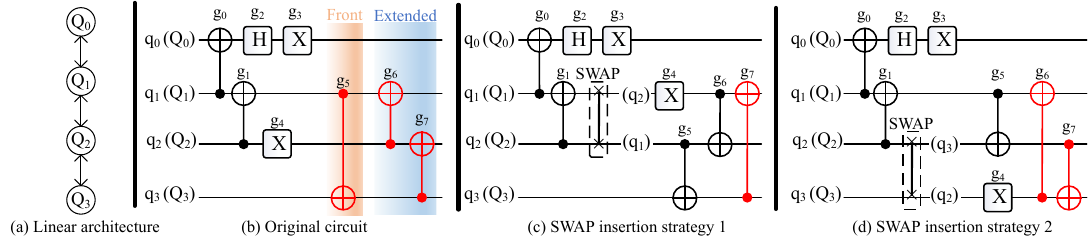}
  \caption{The basic initial mapping relationship and the two given strategies for inserting SWAP gates}
  \label{quantumcircuits}
\end{figure*}

The quantum circuit is typically represented as \( LC(q_l, g) \), where \( q_l \) denotes the logical qubits and \( g \) represents the gates. Each line in the circuit diagram represents a qubit, with quantum gates ordered left to right by execution. For example, in Fig. \ref{quantumcircuits}(b), the quantum circuit consists of 4 qubits and 8 quantum gates, with single-qubit gates like $X$ and $H$, and two-qubit gates like $g_0$ and $g_5$, which are represented as \( {CX}(q_1, q_0) \) and \( {CX}(q_1, q_3) \), respectively. Finally, after all gates are executed, the result of the circuit is obtained through measurements at the end of the quantum circuit.

% However, the above circuit represents a logical quantum circuit and does not account for the connectivity constraints between qubits in actual quantum devices.

\subsection{Qubit Mapping and Routing Problem}
A quantum circuit $LC(q_l, g)$ and quantum device architecture $AG(Q_p, E)$ are considered, restricted to single-qubit and CNOT gates for simplicity.

The qubit mapping problem involves establishing an initial one-to-one correspondence between logical qubits and physical qubits, denoted as $\pi:\{q_l \rightarrow Q_p \}$. This initial mapping can be either randomly assigned or determined by a simple heuristic. As illustrated in Fig. \ref{quantumcircuits}(b), the mapping is represented by $\pi^{init}:\{q_i \rightarrow Q_i \}$, which satisfies the architectural constraints shown in Fig. \ref{quantumcircuits}(a) for certain CNOT gates (e.g., $g_0$ and $g_1$) within the circuit.

The routing problem is defined as follows: Given an initial mapping, certain two-qubit gates in the circuit may involve logical qubits mapped to physically distant qubits. To address this, SWAP gates are dynamically inserted to adjust the mapping, ensuring that the logical qubits in a two-qubit gate are mapped to directly connected physical qubits, enabling gate execution. As shown in Fig. \ref{quantumcircuits}(b), the gates marked in red ($g_5$, $g_6$, $g_7$) cannot be executed under the mapping \( \pi^{init} \) and the gate dependency constraints, two feasible SWAP insertion strategies are illustrated in Fig. \ref{quantumcircuits}(c) or Fig. \ref{quantumcircuits}(d). 
% This adjustment ensures that the two physical qubits involved in the relevant gate are directly connected, allowing the gate to be executed. By continuously inserting SWAP gates, a quantum circuit that satisfies the constraints of the given physical architecture is ultimately formed.

% However, inserting SWAP gates not only increases the gate count and circuit depth but also introduces additional noise, significantly impacting the success rate of quantum algorithms. Therefore, the primary goal of existing quantum circuit mapping algorithms is to minimize the number of SWAP gates or the circuit depth, as both factors critically affect the fidelity and execution time of quantum circuits.
\subsection{Circuit Depth and Single Qubit Gate Storage}
To effectively track and optimize the circuit depth for each qubit in a quantum circuit, the $DP$ is introduced. Specifically, \( DP(Q) \) represents the current depth of a physical qubit \( Q \), initialized to 0. For single-qubit gates, the \( DP \) value of the corresponding node is incremented by 1. For two-qubit gates acting on \( q_0 \) and \( q_1 \), the depths of \( Q_{q_0} \) and \( Q_{q_1} \) are updated as \(DP(Q_{q_0}) = DP(Q_{q_1}) = \max(DP(Q_{q_0}), DP(Q_{q_1})) + 1 \). 
For SWAP gates, the update rule is \(DP(Q_{q_1}) = DP(Q_{q_0}) = \max(DP(Q_{q_1}), DP(Q_{q_0})) + 3 \), reflecting their decomposition into three CNOT gates.

In addition, the execution order of single-qubit gates impacts circuit depth, making their storage in $SG$ essential. During circuit execution, all encountered single-qubit gates are stored in $SG$. When a two-qubit gate \((q_0, q_1)\) satisfying device connectivity constraints is met, both the single-qubit gates stored in \(SG(Q_{q_0})\) and \(SG(Q_{q_1})\) will be popped from $SG$ and applied to the circuit. For a two-qubit gate that does not satisfy the architecture constraints, SWAP insertion is preceded by computing \(s = \arg\min(DP(Q_{q_0}), DP(Q_{q_1}))\) and removing \(k = \min \left( |DP(Q_{q_0}) - DP(Q_{q_1})|, len(SG(s)) \right)\) single-qubit gates from $SG(s)$ to balance $DP$ values. 

\section{Our method}\label{sec3}
This section primarily presents the proposed TANGO method. First, we introduce the dual-factor optimization for initial mapping, which improves mapping quality by balancing the impact of qubit placement on mapped and unmapped nodes. Next, a novel two-stage qubit routing optimization method is proposed, which considers factors such as the number of executable gates, circuit depth, and a bidirectional SWAP look strategy, thereby enabling the co-optimization of gate count and circuit depth.

\subsection{Dual-Factor Optimization for Initial Mapping}
The objective of the initial mapping is to determine an optimal placement of qubits on a connectivity-constrained physical architecture, minimizing SWAP gates inserted during the routing process, thereby reducing the overall gate count and circuit depth. To achieve this, we propose a dual-factor evaluation function to enhance the initial mapping process. First, gate weights within the circuit should be defined, as they determine the mapping priority of the qubits.

Naturally, for a quantum circuit \( LC(q_l, g) \), gate weights will gradually decrease due to gate dependency, as it directly impacts the execution of subsequent gates. While some existing methods \cite{qian2023method, chang2021mapping} assign a unique and monotonically decreasing weight to each gate,  they fail to adequately account for the weight distribution of parallel gates within the circuit. This insufficient consideration of weight information may adversely impact the effectiveness of the initial mapping. Therefore, we define the weights based on the parallelism rules of gates in the circuit. Inspired by reward function design in reinforcement learning, a decremental form is employed to define the weight \( W_g(g_i) \) for each gate, specifically represented as:
\begin{equation}
    W_g(g_i) = \gamma^{{layer}_i},
\end{equation}
where $\gamma$ indicates the initial weight, \(g_i\) represents the \(i\)-th gate in the circuit, and \({layer}_i\) denotes the layer of the quantum circuit where \(g_i\) is located.

Once the weight for each quantum gate is defined, the priority of each qubit is determined by the sum of the weights of the gates involving the current qubit. The weight for each qubit is defined as:
\begin{equation}
    W_q(q_i) = \sum_{g_j} W_g(g_j), \quad \text{where} \quad q_i \in g_j.
\end{equation}
Similarly, for the weight definition between two qubits, it is given by the sum of the gate weights involving both qubits, expressed as:
\begin{equation}
    e_{qw} = \sum_i W_g(g_i), \quad \text{where} \quad q\ \text{ and }\ w \in g_i.
\end{equation}
After calculating the weights for all qubits, the qubits with higher weights are prioritized for execution, as their associated gates are more concentrated at the front of the circuit. Specifically, by sorting the qubits based on \(W_q(q_i)\), the qubit with the highest weight is mapped first, followed by qubits with progressively lower weights. The qubit with the highest weights is preferentially mapped to the center point of the quantum device, as this is considered the core position within the architecture. For the remaining logical qubits to be mapped, the algorithm primarily considers two key factors: first, the impact of placing a qubit at a designated location on the already mapped qubits; second, the effect on the unmapped qubits. By considering both factors, the overall impact of qubit placement can be comprehensively evaluated.

To analyze the first factor, we design a scoring function to assess the impact on mapped nodes:
\begin{equation}
    f_p(q, Q_v) = \sum_w \left(1 - \frac{{dis}(Q_v, Q_w)}{{dia}(AG)} \right) * e_{qw},
\end{equation}
where \(q\) is the qubit node to be mapped to the quantum device, \(Q_v\) represents the corresponding position on the quantum device, \(w\) is a previously placed logical neighbor of \(q\), \({dia}(AG)\) is the diameter of the architecture \(AG\), and \({dis}(Q_v, Q_w)\) denotes the distance between these two nodes on the quantum device. The motivation behind this function is to place \(q\) as close as possible to its logical neighbor \(w\) and to ensure that the weight between the two nodes is larger, thereby maximizing the value of the objective function \(f_p\) and potentially increasing the likelihood of executing subsequent gates.

The second factor involves considering the unplaced logical neighbors of qubit \(q\), for which we define the function \(f_{{up}}\) as follows:
\begin{equation}
    f_{{up}}(q, Q_v) = \frac{(f(Q_v) - |q'|) * \sum_u e_{qu}}{\max(f(Q_v), |q'|)},
\end{equation}
where \(u\) represents an unplaced logical neighbor of \(q\), \(|q'|\) denotes the number of unmapped logical qubits connected to \(q\), and \(f(Q_v)\) represents the number of available neighboring nodes at position \(Q_v\) on the quantum device. Similarly, the motivation behind this function is to identify suitable free positions for placing \( q \)’s logical neighbors, prioritizing qubits with higher weights. If the available positions around $Q_v$ are insufficient to accommodate all of $q$'s neighbors, which results in the function value being negative, effectively acting as a penalty term. 

During the calculation of both scoring functions, the values of 
$f_p$ and \(f_{up}\) may vary significantly. To address this, both functions are standardized using $Z-scores$ to bring them to the same scale, resulting in $f_{p_z}$ and $f_{up_z}$. Furthermore, adjustable parameters \(\alpha\) and \(\beta\)  are introduced to appropriately weight these functions, thereby achieving a balanced relationship between them. The final scoring function is defined as:
\begin{equation}
  F (q)= \arg \max\limits_{Q_v} \left( \alpha * f_{p_z}(q, Q_v) + \beta * f_{up_z}(q, Q_v) \right).  
\end{equation}
Based on this approach, the placement position for each qubit \(q\) in the quantum architecture is determined by selecting the position \(Q_v\) that maximizes the function value \(F\). This process is iteratively applied to all qubits in the logical quantum circuit until all logical qubits are successfully mapped to physical nodes in the quantum device. Ultimately, a mapping relationship $\pi^{init}$ is established to guide subsequent qubit routing.

\begin{figure}[t]
  \centering
  \includegraphics{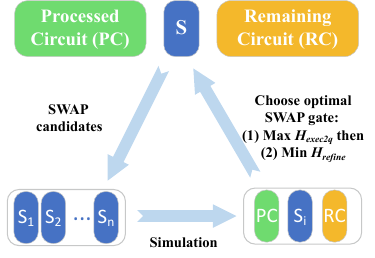}
  \caption{The workflow of the two-stage routing algorithm for searching SWAP gates}
  \label{overallrouting}
\end{figure}

\subsection{Two-Stage Qubit Routing Optimization Algorithm}
After obtaining the initial mapping, additional SWAP gates are often needed to modify the qubit mapping, ensuring the execution of two-qubit gates that may not satisfy the connectivity constraint. To optimize both the number of additional gates and the circuit depth introduced during the mapping process, we propose an innovative two-stage routing algorithm. This algorithm uses the executable gate count as the backbone evaluation metric, with circuit-related features serving as auxiliary decision factors to effectively identify efficient SWAP gates. The workflow is illustrated in Fig. \ref{overallrouting}.

Initially, the SWAP candidates is defined as SWAP\(_{candidate}\) = \{SWAP\(_i\)\}, where each SWAP\(_i\) involves at least one qubit that appears in the first \( k \) two-qubit gates of the remaining circuit. This strategy ensures that the size of the SWAP candidate set never exceeds \( |EG| \), the original search space size determined by the total number of edges in the architecture, thereby effectively reducing the computational resource requirements and enhancing the algorithm's efficiency.

In the first stage, we define \(H_{exec2q} \) as a scoring function to evaluate each SWAP gate within the SWAP$_{candidate}$. Most existing heuristic methods for SWAP gate selection typically rely on the distance between subsequent gates after the mapping transformation. However, we find that the strategy based on the number of executable gates directly addresses the mapping problem more effectively, as opposed to indirectly assessing gate distances to determine the insertion of SWAP gates. This \(H_{exec2q} \) function primarily focuses on maximizing the number of two-qubit gates that can be executed by a single SWAP gate, thereby ensuring that the current SWAP gate is optimal within the local search space. Consequently, the function \(H_{exec2q} \) can be defined as follows:
\begin{equation}\label{exe2q}
H_{exec2q}(SWAP_i) = {Execute}(Gates_{2q}),
\end{equation}
where \( {Execute}(Gates_{2q}) \) represents the number of two-qubit gates that can be executed in the quantum circuit after inserting the SWAP$_i$ gate. A higher \(H_{exec2q} \) value indicates that the current SWAP gate yields a greater benefit within the local scope, as more two-qubit gates can be executed. By selecting {SWAP}\(_i\) from the SWAP candidate set and calculating its corresponding function based on Eq. \ref{exe2q}, the maximum value \( MES \) is obtained as:
\(
MES = \max(H_{exec2q})
\). When \( MES = 0 \), it indicates that no SWAP gate can enable the execution of any two-qubit gates. If \( MES \neq 0 \), multiple SWAP gates may yield the same value. In such cases, further refinement is required to select the optimal SWAP gate by considering additional factors, such as gate distance and circuit depth, which constitutes the objective of the second stage.

For the scenario where \( MES \neq 0 \), motivated by the fact that the gate distance, circuit depth, and routing-based circuit optimization help guide the insertion of SWAP gates, thereby optimizing both the quantum circuit depth and the number of additional gates. Based on this, we define the following refined scoring function \( H_{refine} \), with the objective of minimizing its value:
\begin{equation}
    H_{refine} = H_{decay} + {DP_{max}} - {Reward},
\end{equation}
where $H_{decay}$ highlights the impact of SWAP insertion on the distance between the remaining gates in the circuit and is expressed as:
\begin{equation}
\begin{split}
 H_{decay} & = \max({decay}(Q_i), {decay}(Q_j)) \\ & *\left(\frac{1}{|F|} {Dis}(\pi, F) + w * \frac{1}{|E|} {Dis}(\pi, E)\right) .
\end{split}
\end{equation}
Here, the term “decay” refers to an increase in the decay value \( \delta \) when a SWAP gate is executed between qubits \( Q_i \) and \( Q_j \), where \( \delta \) ranges from \([0, 1]\). This mechanism encourages the algorithm to prioritize qubits that have not recently participated in a SWAP gate, thereby effectively reducing circuit depth. 

\begin{figure}[t]
  \centering
  \includegraphics[width=\columnwidth]{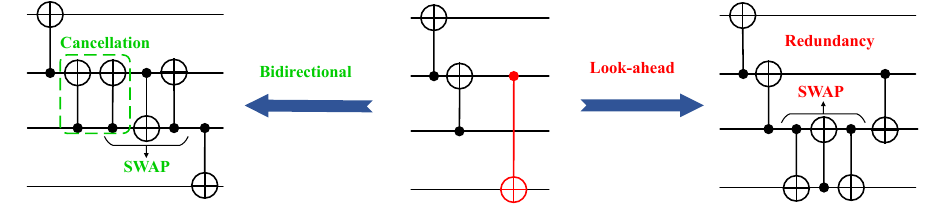}
  \caption{An example of the motivation for the bidirectional  SWAP gate search}
  \label{reward}
\end{figure}
\begin{figure}[t]
  \centering
  \includegraphics[width=\columnwidth]{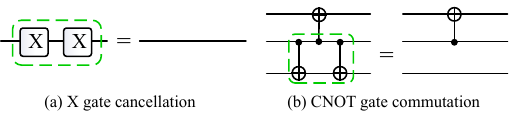}
  \caption{Rules for gate cancellation and commutation of X and CNOT gates}
  \label{CCrule}
\end{figure}
\begin{figure*}[ht]
  \centering
\includegraphics[width=2.12\columnwidth]{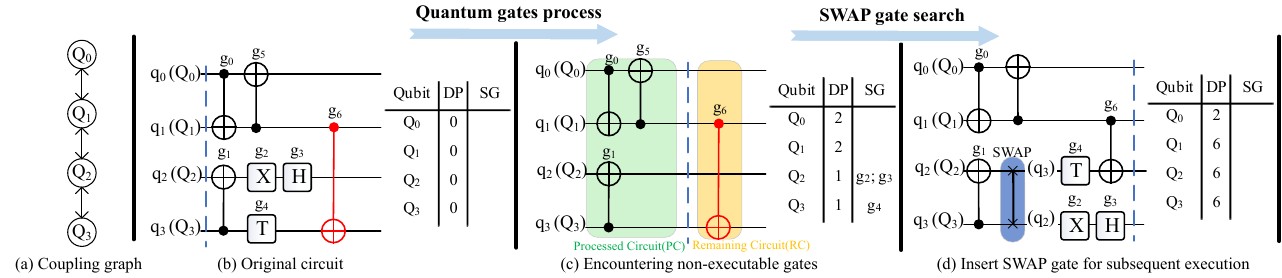}
  \caption{The example of two-stage qubit routing algorithms within a circuit}
  \label{routing}
\end{figure*}The impact of a SWAP gate on subsequent quantum gates is categorized into two types of gate distance contributions: \( F \), representing the front layer, and \( E \), representing the extended layer, are both shown in Fig. \ref{quantumcircuits}(b). The range of \( E \) is usually determined by a predefined look-ahead window, which specifies the number of gates considered within the extended layers. Besides, a discount factor is applied to the gates in the extended layer, as the gates in the front layer \( F \) need to be executed first to enable the execution of subsequent dependent operations.

As discussed in the background section, to optimize circuit depth during the mapping process, we introduce a variable \( DP \) to track the depth of each qubit. Specifically, each time a single-qubit gate is applied to a qubit, its corresponding \( DP \) value increases by 1, reflecting the qubit's current depth within the circuit. Additionally, we define a set \( SG \) to store the single-qubit gates for a given qubit. If the CNOT gates are executable, these single-qubit gates are applied to the relevant qubits accordingly.

The term \({DP_{max}}\), defined as \( {DP_{max}} = \max(DP_i) \), indicates the maximum depth of the circuit after inserting a SWAP gate at the current position, aiming to reduce the insertion of SWAP gates into qubits with higher depths. 

Moreover, the term $Reward$ is introduced as a novel bidirectional SWAP marker, with its motivation illustrated in the Fig. \ref{reward}. Unlike conventional look-ahead approaches that focus solely on the impact of SWAP insertion on remaining gates, we also consider its influence on preceding gates, forming a bidirectional look approach. This is because a SWAP operation may facilitate optimization with preceding gates, thereby reducing both the number of gates and the overall circuit depth. Initially set to 0, $Reward$ is updated based on a tracking set \( REC \), which records the most recent gate executed for each qubit. If \( {REC}[Q_i] = {REC}[Q_j] \), this implies that \( Q_i \) and \( Q_j \) share the same gate. In this case, a positive $Reward$ is assigned, reflecting the potential for the SWAP gate to cancel out the two-qubit gates preceding it, further simplifying the quantum circuit. Finally, by minimizing the function \( H_{refine} \), we determine the optimal SWAP gate for insertion, enabling the execution of subsequent gates.

For the second scenario, where \( MES = 0 \), the process becomes relatively straightforward. The shortest-distance CNOT gate from the front layer that cannot be executed is identified, and a SWAP gate is applied to reduce its distance by 1. Simultaneously, it is ensured that this SWAP gate satisfies the \( \min(H_{decay}) \) condition among the available candidates. This strategy enables the algorithm to efficiently execute high-priority gates while minimizing the distances of subsequent gates, hence optimizing the overall performance of the quantum circuit.

By repeatedly performing this routing process, all quantum gates in the circuit are eventually executed. Similarly to the SABRE algorithm's initial mapping generation phase, bidirectional traversal techniques are applied. For the entire unmapped quantum circuit, denoted as the Original Circuit (OC), the initial mapping and routing algorithm is applied to make the OC executable on the architecture. The OC is then reversed to obtain \(\rm OC_{RS}\), and the mapping relationship generated at the end of the previous mapped circuit is input into  \(\rm OC_{RS}\). After completing the routing phase and executing all quantum gates, the circuit is reversed again to obtain the OC, and the routing process is performed again. Finally, we evaluate the minimum circuit depth achieved over these three runs. It is important to clarify that the focus of our apporach's rerunning lies in the circuit depth or total gate count obtained during the bidirectional traversals, rather than merely for generating a good initial mapping. This insight helps prevent the significant computational resource waste that would occur within the SABRE algorithm.

To further enhance the performance of the mapped quantum circuit, we have integrated relevant circuit optimization strategies into our algorithm. As illustrated in Fig. \ref{CCrule}, gate cancellation and commutation and techniques are applied, which play a crucial role in significantly optimizing the quantum circuit. Specifically, for multiple single-qubit gates, such as two adjacent X gates, they can be directly canceled. Similarly, optimization rules can be applied to two-qubit gates, particularly CNOT gates, where consecutive gates sharing the same control or target qubits can often be canceled or commuted. What's more, the adaptive decomposition of SWAP gates is determined based on the gate directions preserved in the tracking set $REC$, allowing the adaptive strategy to eliminate CNOT gates. It was found that the SWAP gate in the SQGM algorithm only has a single decomposition form and lacks this adaptive approach, which is critical for quantum circuit optimization. These optimizations effectively simplify the quantum circuit while maintaining functional equivalence, thus reducing the gate counts and the circuit depth. 

To better understand our proposed two-stage routing optimization algorithm, we present an illustrative example of the overall routing process in \ref{routing}. Fig. \ref{routing}(b) presents the initial mapping relationship: 
\begin{equation}
 \pi^{init} = \{q_0 : Q_0, q_1 : Q_1, q_2 : Q_2, q_3 : Q_3\}.
\end{equation}
Initially, $DP$ is set to 0, and \( SG \) is initialized as null. During quantum circuit execution, single-qubit gates are stored in \( SG \) until an unexecutable CNOT gate, \( g_6 \), is encountered. The decision for SWAP insertion is based on the calculation results from Eq. \ref{exe2q}. Specifically, 
\begin{equation} {H_{exec2q}(SWAP}(q_2, q_3){)} = {H_{exec2q}(SWAP}(q_1, q_2){)}.
\end{equation}
In the second stage, these two candidate SWAP gates are further evaluated to determine the optimal choice. It is observed that \( {DP_{max}(SWAP}(q_1, q_2){)} = 5 \) is greater than \( {DP_{max}(SWAP}(q_2, q_3){)} = 4 \). Besides, for $SWAP((q_2, q_3))$ gate, since \( {REC}[Q_2] = {REC}[Q_3] \), the corresponding $Reward$ value is positive. Given that the \( H_{decay} \) values are equal for both SWAP gates, the refinement function yields
\begin{equation}
 H_{refine}({SWAP}(q_2, q_3)) < H_{refine}({SWAP}(q_1, q_2)),
\end{equation}
indicating that $SWAP((q_2, q_3))$ is the optimal choice. Notably, this choice introduces a smaller circuit depth compared to ${SWAP}(q_1, q_2)$.

\begin{figure}[t]
  \centering
  \includegraphics[width=\columnwidth]{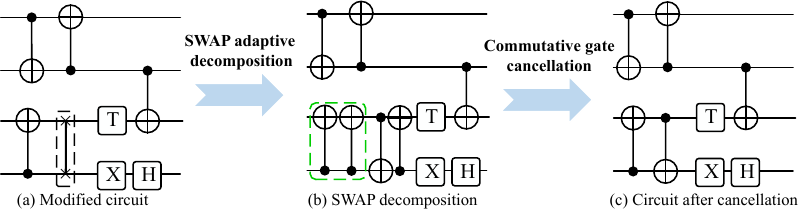}
  \caption{Quantum circuit optimized through gate cancellation and commutation}
  \label{cancellation}
\end{figure}

Additionally, the effectiveness of the circuit optimization strategy is demonstrated in Fig. \ref{cancellation}. After SWAP gate decomposition and optimization, the quantum circuit in Fig. \ref{cancellation}(c) shows a significant reduction in both the number of gates and circuit depth compared to the original circuit in Fig. \ref{cancellation}(a). Specifically, the original circuit contains 7 CNOT gates with a depth of 6, whereas the optimized circuit includes only 5 CNOT gates and a reduced depth of 4. These results effectively demonstrate that the proposed mapping algorithm, through the integration of SWAP adaptive decomposition and commutative gate cancellation, significantly reduces both the depth and gate count of the quantum circuit, yielding notable optimization results. 

\subsection{Complexity Analysis}

For a given quantum circuit \( LC(q_l, g) \) and a quantum architecture graph \( AG(Q_p, E) \), the time complexity analysis is as follows:

During the initial mapping phase, suppose there are \( O(|q_l|) \) logical qubits, each of which can choose from \( O(|Q_p|) \) physical qubit locations. In the worst case, the time complexity for evaluating each possible choice is \( O(|q_l|) \). Since the initial mapping involves determining the positions for all logical qubits, the overall time complexity for this phase is \( O(|q_l|^2 \cdot |Q_p|) \).

In the intermediate qubit routing phase, the candidate set size for each SWAP operation is \( O(|E|) \), where \( |E| \) represents the number of edges in the architecture graph. Assuming each evaluation enables only one gate operation, \( O(|g|) \) evaluations are required, where \( |g| \) is the total number of gates in the quantum circuit. Additionally, the computational complexity of the heuristic function is \( O(|g|) \). Therefore, the time complexity of the routing phase is \( O(|g|^2 \cdot |E|) \).

Combining the complexities of the initial mapping and intermediate routing phases, the overall time complexity of the algorithm is:
\begin{equation}
  O(|q_l|^2 \cdot |Q_p| + |g|^2 \cdot |E|).  
\end{equation}

This complexity indicates that the algorithm's dependence on the input parameters is polynomial in nature, ensuring good scalability and computational efficiency. This is further validated by subsequent experiments, where the algorithm demonstrates reasonable runtime when handling large-scale quantum circuits.

\section{Experiments and Evaluation}\label{sec4}
To comprehensively demonstrate the effectiveness of the quantum circuit mapping algorithm proposed in this paper, its performance was assessed across different circuit dataset sizes and architectures with varying connectivity sparsity. 
%Several widely-used quantum circuit datasets were selected for experimentation. These circuits contain no more than 20 qubits, and the experiments were conducted on the IBM Q{20} architecture. For circuits with qubit numbers ranging from 20 to 40, the relatively sparsely connected Rochester architecture was selected as the benchmark. Given its frequent practical use, assessing its performance is of considerable significance.
\subsection{Methodology}

\textbf{Dataset:} We extracted common quantum circuits from existing benchmark tests as representative circuits for execution on the IBM Q20 architecture. These circuits consist solely of two-qubit gates (CX gates) and single-qubit gates (H, T, T$^{\dag}$, etc.). Additionally, to assess the algorithm's performance on large-scale quantum circuit sets, we selected quantum circuits used in the field of chemistry. These circuits vary in gate count, ranging from tens of thousands to hundreds of thousands, and include only two-qubit gates (CX) and some single-qubit gates (U$_1$, U$_3$), with the number of qubits ranging from 20 to 40. These datasets provide a comprehensive evaluation of the algorithm’s performance under different scenarios.

\textbf{Hardware Model:} This study is primarily evaluated on two common quantum computing architectures. For circuits with fewer than 20 qubits, the IBM Q20 architecture, featuring 20 qubits and relatively dense connectivity, was chosen as the benchmark. For circuits with more qubits (ranging from 20 to 40), experiments were conducted on the Rochester architecture, which is configured with 53 qubits and relatively sparse connectivity.

\textbf{Algorithm Configuration:} During algorithm execution, the parameters for the initial mapping phase, \( \alpha \) and \( \beta \), were each set to 0.5, and \(\gamma\) was set to 0.99. Besides, \( \delta \) was set to 0.001, the positive $Reward$ was set to 0.5, and the weight parameter \( w \) was also set to 0.5. Each algorithm was run three times, and the best result was selected for comparison.

\textbf{Experimental Platform:} All experiments were conducted on a personal computer with the following configuration: Intel(R) Core(TM) i5-8300H CPU @ 2.30GHz and 16GB DDR4 memory.

\textbf{Evaluation Metrics:} Under various architectures and datasets, several common metrics were compared, including the algorithm's runtime, the final number of two-qubit gates in the circuit, and the overall circuit depth. These factors are crucial for evaluating the performance of quantum algorithms.

\textbf{Comparative Algorithms:} We compared the proposed quantum circuit mapping algorithm with the NASSC\cite{liu2022not} algorithm and SQGM\cite{sqgm2023single} algorithm, which is designed for optimizing circuit depth. It is important to note that both NASSC and SQGM use the SABRE algorithm’s initial mapping method during their initial mapping phase and have made significant contributions to the academic community.

\subsection{Comparison with NASSC and SQGM on the IBM Q20 Architecture}

\begin{table*}[t]
    \centering
    \caption{Comparison of CNOT count and circuit depth between NASSC and SQGM with TANGO on the IBM Q20 for 'small' circuits}	
    \label{SMALL}
    \begin{threeparttable}
    \resizebox{\linewidth}{!}{
    \begin{tabular}{ccccccccccccccccc}
    \hline
        \multirow{2}*{Benchmark name}  & \multirow{2}*{$\#qubits$} & \multirow{2}*{$CNOT_{total}$} & \multirow{2}*{$Depth_{total}$} & ~ & NASSC & ~ & ~ & SQGM & ~ & ~ & TANGO & ~ & ~ & \multicolumn{2}{c}{Comparison} & ~ \\ 
        ~ & ~ & ~ & ~ & $g_1$ & $d_1$ & $RT$ & $g_2$ & $d_2$ & $RT$ & $g_3$ & $d_3$ & $RT$ & $\Delta_{g_1}$ & $\Delta_{d_1}$ & $\Delta_{g_2}$ & $\Delta_{d_2}$ \\ \hline
3\_17\_13.qasm & 3 & 17 & 22 & 33 & 41 & 0.1 & 36 & 41 & 0.05 & 17 & 22 & 0.06  & 48.48\% & 46.34\% & 52.78\% & 46.34\% \\ 
        miller\_11.qasm & 3 & 23 & 29 & 39 & 50 & 0.13 & 43 & 47 & 0.06 & 23 & 29 & 0.08  & 41.03\% & 42.00\% & 46.51\% & 38.30\% \\ 
        decod24-v0\_38.qasm & 4 & 23 & 30 & 27 & 35 & 0.1 & 30 & 37 & 0.05 & 21 & 28 & 0.12  & 22.22\% & 20.00\% & 30.00\% & 24.32\% \\ 
        rd32-v1\_68.qasm & 4 & 16 & 21 & 30 & 38 & 0.1 & 19 & 25 & 0.04 & 12 & 19 & 0.04  & 60.00\% & 50.00\% & 36.84\% & 24.00\% \\ 
        4gt13\_91.qasm & 5 & 49 & 61 & 74 & 87 & 0.19 & 99 & 95 & 0.09 & 55 & 66 & 0.16  & 25.68\% & 24.14\% & 44.44\% & 30.53\% \\ 
        4gt5\_75.qasm & 5 & 38 & 47 & 54 & 63 & 0.14 & 63 & 73 & 0.14 & 44 & 54 & 0.11  & 18.52\% & 14.29\% & 30.16\% & 26.03\% \\ 
        4mod5-v0\_19.qasm & 5 & 16 & 21 & 22 & 27 & 0.08 & 29 & 30 & 0.04 & 16 & 21 & 0.06  & 27.27\% & 22.22\% & 44.83\% & 30.00\% \\ 
        alu-v3\_34.qasm & 5 & 24 & 30 & 54 & 62 & 0.15 & 55 & 60 & 0.06 & 27 & 33 & 0.07  & 50.00\% & 46.77\% & 50.91\% & 45.00\% \\ 
        alu-v4\_36.qasm & 5 & 51 & 66 & 80 & 89 & 0.25 & 82 & 87 & 0.1 & 55 & 70 & 0.11  & 31.25\% & 21.35\% & 32.93\% & 19.54\% \\ 
        decod24-v3\_45.qasm & 5 & 64 & 84 & 99 & 115 & 0.28 & 174 & 163 & 0.17 & 81 & 98 & 0.43  & 18.18\% & 14.78\% & 53.45\% & 39.88\% \\ 
        mini-alu\_167.qasm & 5 & 126 & 162 & 201 & 234 & 0.49 & 205 & 225 & 0.22 & 182 & 209 & 0.37  & 9.45\% & 10.68\% & 11.22\% & 7.11\% \\ 
        mod5d1\_63.qasm & 5 & 13 & 13 & 25 & 27 & 0.07 & 13 & 13 & 0.03 & 13 & 13 & 0.04  & 48.00\% & 51.85\% & 0.00\% & 0.00\% \\ 
        mod5mils\_65.qasm & 5 & 16 & 21 & 25 & 31 & 0.09 & 25 & 28 & 0.04 & 16 & 21 & 0.05  & 36.00\% & 32.26\% & 36.00\% & 25.00\% \\ 
        one-two-three-v0\_98.qasm & 5 & 65 & 82 & 115 & 137 & 0.3 & 139 & 137 & 0.14 & 101 & 114 & 0.21  & 12.17\% & 16.79\% & 27.34\% & 16.79\% \\ 
        one-two-three-v1\_99.qasm & 5 & 59 & 76 & 127 & 136 & 0.32 & 101 & 103 & 0.1 & 107 & 111 & 0.26  & 15.75\% & 18.38\% & -5.94\% & -7.77\% \\ 
        one-two-three-v3\_101.qasm & 5 & 32 & 40 & 44 & 53 & 0.16 & 42 & 50 & 0.06 & 42 & 53 & 0.11  & 4.55\% & 0.00\% & 0.00\% & -6.00\% \\ 
        qe\_qft\_4.qasm & 5 & 27 & 36 & 30 & 38 & 0.11 & 33 & 41 & 0.07 & 27 & 36 & 0.10  & 10.00\% & 5.26\% & 18.18\% & 12.20\% \\ 
        rd32\_270.qasm & 5 & 36 & 47 & 57 & 68 & 0.16 & 83 & 85 & 0.09 & 44 & 55 & 0.11  & 22.81\% & 19.12\% & 46.99\% & 35.29\% \\ 
        \textbf{Geometric mean} & ~ & ~ & ~ & ~ & ~ & ~ & ~ & ~ & ~ & ~ & ~ & ~ & \textbf{27.85\%} & \textbf{25.35\%} & \textbf{30.92\%} & \textbf{22.59\%} \\ \hline
    \end{tabular}
    }
         \begin{tablenotes}    %这行要添加， 从这开始
        \footnotesize            %这行要添加
\item[*] {Note:\( g_1, g_2, g_3 \) and \( d_1, d_2, d_3 \) represent the total number of CNOT gates and circuit depth after mapping with NASSC, SQGM, and TANGO algorithms,

respectively. RT refers to the runtime (s). \( \Delta_{g_i} = 1 - \frac{g_3}{g_i} \), \( \Delta_{d_i} = 1 - \frac{d_3}{d_i} \).} 
      \end{tablenotes}            %这行要添加
    \end{threeparttable}       %这行要添加，到这里结束
\end{table*}

\begin{table*}[t]
    \centering
    \caption{Comparison of CNOT count and circuit depth between NASSC and SQGM with TANGO on the IBM  Q20 for 'medium' circuits}
    \label{MIDDLE}
    \resizebox{\linewidth}{!}{

    \begin{tabular}{ccccccccccccccccc}
    \hline
        \multirow{2}*{Benchmark name}  & \multirow{2}*{$\#qubits$} & \multirow{2}*{$CNOT_{total}$} & \multirow{2}*{$Depth_{total}$} & ~ & NASSC & ~ & ~ & SQGM & ~ & ~ & TANGO & ~ & ~ & \multicolumn{2}{c}{Comparison} & ~ \\ 
        ~ & ~ & ~ & ~ & $g_1$ & $d_1$ & $RT$ & $g_2$ & $d_2$ & $RT$ & $g_3$ & $d_3$ & $RT$ & $\Delta_{g_1}$ & $\Delta_{d_1}$ & $\Delta_{g_2}$ & $\Delta_{d_2}$ \\ \hline
 4gt12-v0\_87.qasm & 6 & 112 & 131 & 218 & 249 & 0.52 & 263 & 227 & 0.23 & 189 & 191 & 0.56  & 13.30\% & 23.29\% & 28.14\% & 15.86\% \\ 
        4gt4-v0\_78.qasm & 6 & 109 & 137 & 137 & 158 & 0.32 & 195 & 200 & 0.25 & 120 & 145 & 0.22  & 12.41\% & 8.23\% & 38.46\% & 27.50\% \\ 
        4gt4-v0\_79.qasm & 6 & 105 & 132 & 175 & 203 & 0.49 & 189 & 203 & 0.18 & 115 & 141 & 0.21  & 34.29\% & 30.54\% & 39.15\% & 30.54\% \\ 
        4gt4-v0\_80.qasm & 6 & 79 & 101 & 124 & 144 & 0.29 & 145 & 143 & 0.13 & 100 & 119 & 0.31  & 19.35\% & 17.36\% & 31.03\% & 16.78\% \\ 
        4gt4-v1\_74.qasm & 6 & 119 & 154 & 211 & 255 & 0.54 & 277 & 266 & 0.26 & 198 & 209 & 0.44  & 6.16\% & 18.04\% & 28.52\% & 21.43\% \\
        decod24-enable\_126.qasm & 6 & 149 & 190 & 246 & 291 & 0.63 & 330 & 317 & 0.26 & 210 & 243 & 0.66  & 14.63\% & 16.49\% & 36.36\% & 23.34\% \\
        ex3\_229.qasm & 6 & 175 & 226 & 353 & 406 & 0.82 & 352 & 352 & 0.3 & 230 & 281 & 0.44  & 34.84\% & 30.79\% & 34.66\% & 20.17\% \\
        4mod5-bdd\_287.qasm & 7 & 31 & 41 & 68 & 75 & 0.2 & 52 & 60 & 0.06 & 49 & 58 & 0.15  & 27.94\% & 22.67\% & 5.77\% & 3.33\% \\
        C17\_204.qasm & 7 & 205 & 253 & 383 & 424 & 0.93 & 392 & 385 & 0.35 & 321 & 349 & 0.82  & 16.19\% & 17.69\% & 18.11\% & 9.35\% \\
        rd53\_130.qasm & 7 & 448 & 569 & 800 & 872 & 1.81 & 996 & 896 & 0.88 & 773 & 834 & 1.95  & 3.37\% & 4.36\% & 22.39\% & 6.92\% \\
        rd53\_131.qasm & 7 & 200 & 261 & 294 & 347 & 0.68 & 338 & 327 & 0.35 & 291 & 325 & 0.77  & 1.02\% & 6.34\% & 13.91\% & 0.61\% \\
        rd53\_135.qasm & 7 & 134 & 159 & 211 & 216 & 0.5 & 287 & 262 & 0.24 & 195 & 209 & 0.46  & 7.58\% & 3.24\% & 32.06\% & 20.23\% \\
        cm82a\_208.qasm & 8 & 283 & 337 & 520 & 571 & 1.33 & 501 & 466 & 0.59 & 406 & 451 & 0.93  & 21.92\% & 21.02\% & 18.96\% & 3.22\% \\
        rd53\_138.qasm & 8 & 60 & 56 & 115 & 118 & 0.3 & 123 & 90 & 0.12 & 89 & 82 & 0.21  & 22.61\% & 30.51\% & 27.64\% & 8.89\% \\
        urf2\_152.qasm & 8 & 35210 & 44100 & 62829 & 70800 & 146.38 & 75982 & 69578 & 85.37 & 59518 & 62905 & 136.64  & 5.27\% & 11.15\% & 21.67\% & 9.59\% \\
        urf2\_277.qasm & 8 & 10066 & 11390 & 20031 & 20399 & 52.52 & 22940 & 18777 & 18.04 & 18073 & 17557 & 47.34  & 9.77\% & 13.93\% & 21.22\% & 6.50\% \\
        con1\_216.qasm & 9 & 415 & 508 & 749 & 816 & 1.86 & 932 & 744 & 0.81 & 699 & 715 & 1.60  & 6.68\% & 12.38\% & 25.00\% & 3.90\% \\
        hwb8\_113.qasm & 9 & 30372 & 38717 & 53181 & 58913 & 140.91 & 67014 & 59969 & 73.27 & 50572 & 53636 & 114.01  & 4.91\% & 8.96\% & 24.54\% & 10.56\% \\
        urf1\_149.qasm & 9 & 80878 & 99585 & 137512 & 154449 & 364.66 & 173067 & 156070 & 198.74 & 133095 & 139792 & 298.18  & 3.21\% & 9.49\% & 23.10\% & 10.43\% \\
        urf1\_278.qasm & 9 & 26692 & 30955 & 48914 & 50696 & 126.27 & 61833 & 51028 & 67.62 & 45578 & 44889 & 124.68  & 6.82\% & 11.45\% & 26.29\% & 12.03\% \\
        urf5\_158.qasm & 9 & 71932 & 89145 & 128754 & 142868 & 316.42 & 154649 & 139324 & 876.05 & 117071 & 124309 & 276.07  & 9.07\% & 12.99\% & 24.30\% & 10.78\% \\
        urf5\_280.qasm & 9 & 23764 & 27822 & 42042 & 44640 & 119.1 & 53690 & 45318 & 45.65 & 40796 & 41344 & 103.33  & 2.96\% & 7.38\% & 24.02\% & 8.77\% \\
        \textbf{Geometric mean} & ~ & ~ & ~ & ~ & ~ & ~ & ~ & ~ & ~ & ~ & ~ & ~ & \textbf{12.92\%} & \textbf{15.38\%} & \textbf{25.69\%} & \textbf{12.76\%} \\ \hline
    \end{tabular}
}        
\end{table*}

To comprehensively evaluate the performance of the algorithm across datasets of varying scales, circuits with five or fewer qubits are categorized as “small”, those with more than six but no more than nine qubits as “medium”, and circuits with ten or more qubits as “large”. The IBM Q20 architecture, one of the most commonly used in quantum algorithms, is widely adopted due to its dense connectivity, which introduces a more complex optimization problem. In the current dataset of circuits, the TANGO demonstrates strong performance and is able to find near-optimal solutions. Compared to the other two heuristic-based mapping algorithms, the presented algorithm effectively reduces both gate count and circuit depth. 

\begin{figure}[t]
  \centering
  \includegraphics[width=\columnwidth]{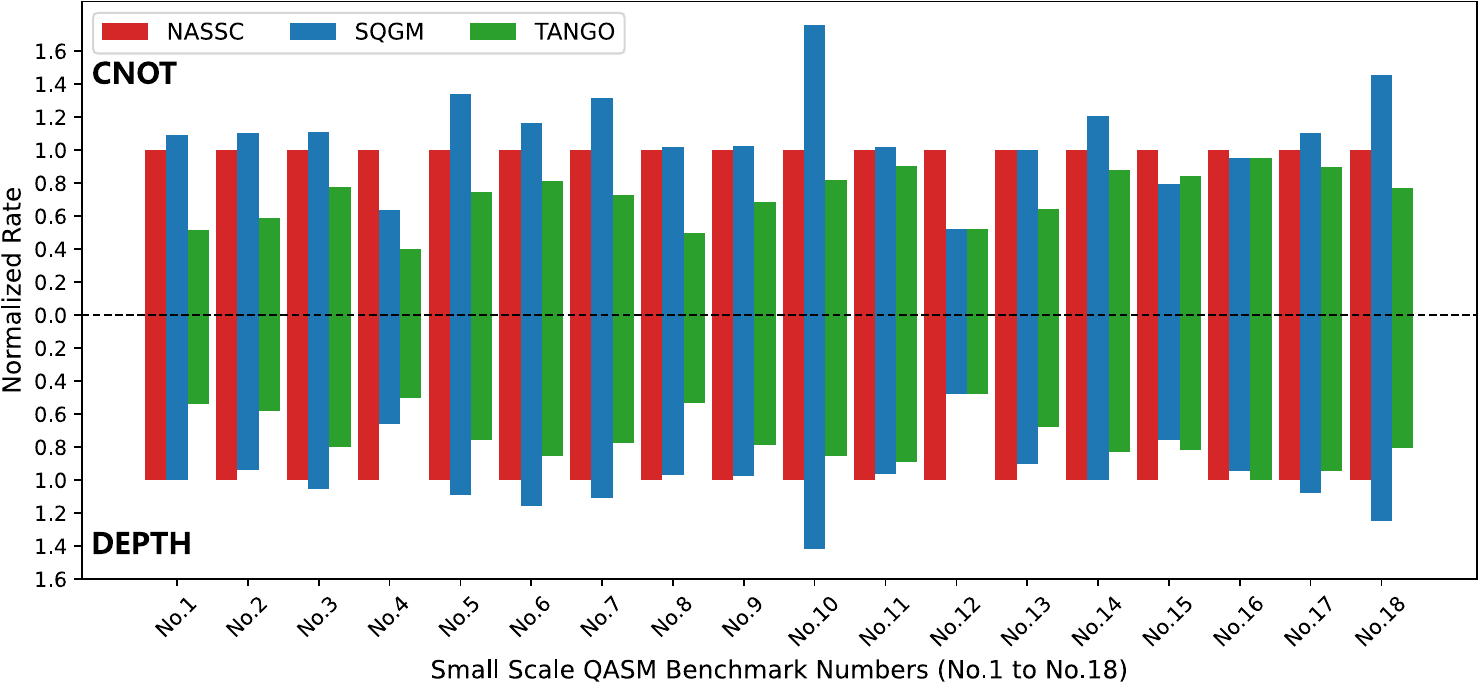}
  \caption{Comparison of total CNOT gates and circuit depth for small-scale quantum circuits on IBM  Q20 (normalize to NASSC)}
  \label{small}
\end{figure}
\begin{figure}[t]
  \centering
  \includegraphics[width=\columnwidth]{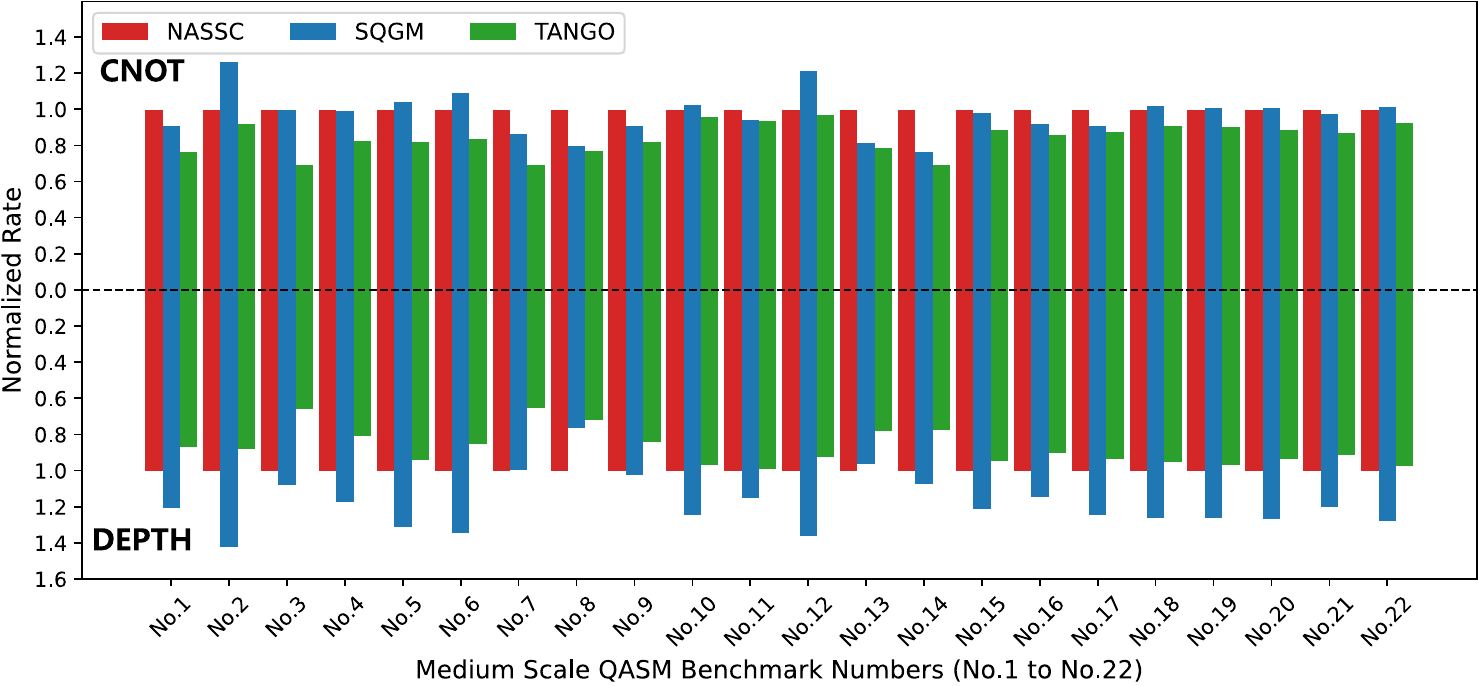}
  \caption{Comparison of total CNOT gates and circuit depth for medium-scale quantum circuits on IBM  Q20 (normalize to NASSC)}
  \label{middle}
\end{figure}

\begin{table*}[!ht]
    \centering
    \caption{Comparison of CNOT count and circuit depth between NASSC and SQGM with TANGO on the IBM Q20 for 'large' circuits}
    \label{LARGE}
    
    \resizebox{\linewidth}{!}{
    \begin{tabular}{ccccccccccccccccc}
    \hline
        \multirow{2}*{Benchmark name}  & \multirow{2}*{$\#qubits$} & \multirow{2}*{$CNOT_{total}$} & \multirow{2}*{$Depth_{total}$} & ~ & NASSC & ~ & ~ & SQGM & ~ & ~ & TANGO & ~ & ~ & \multicolumn{2}{c}{Comparison} & ~ \\ 
        ~ & ~ & ~ & ~ & $g_1$ & $d_1$ & $RT$ & $g_2$ & $d_2$ & $RT$ & $g_3$ & $d_3$ & $RT$ & $\Delta_{g_1}$ & $\Delta_{d_1}$ & $\Delta_{g_2}$ & $\Delta_{d_2}$ \\ \hline
     hwb9\_119.qasm & 10 & 90955 & 116199 & 169675 & 184589 & 437.96 & 205546 & 181239 & 212.26 & 164930 & 168104 & 412.32  & 2.80\% & 8.93\% & 19.76\% & 7.25\% \\ 
        mini\_alu\_305.qasm & 10 & 77 & 69 & 114 & 100 & 0.28 & 125 & 100 & 0.16 & 107 & 91 & 0.31  & 6.14\% & 9.00\% & 14.40\% & 9.00\% \\ 
        qft\_10.qasm & 10 & 90 & 63 & 49 & 38 & 0.31 & 63 & 27 & 0.15 & 36 & 26 & 0.28  & 26.53\% & 31.58\% & 42.86\% & 3.70\% \\ 
        urf3\_155.qasm & 10 & 185276 & 229365 & 338670 & 371971 & 864.05 & 411160 & 365907 & 953.68 & 329215 & 332647 & 781.25  & 2.79\% & 10.57\% & 19.93\% & 9.09\% \\ 
        urf3\_279.qasm & 10 & 60380 & 70702 & 115498 & 118828 & 324.6 & 140545 & 114940 & 655 & 113049 & 107315 & 320.85  & 2.12\% & 9.69\% & 19.56\% & 6.63\% \\ 
        9symml\_195.qasm & 11 & 15232 & 19235 & 27076 & 29351 & 69.52 & 35584 & 30149 & 43.17 & 26067 & 26707 & 64.81  & 3.73\% & 9.01\% & 26.75\% & 11.42\% \\ 
        dc1\_220.qasm & 11 & 833 & 1038 & 1481 & 1622 & 3.6 & 1747 & 1614 & 2.79 & 1194 & 1317 & 2.88  & 19.38\% & 18.80\% & 31.65\% & 18.40\% \\ 
        life\_238.qasm & 11 & 9800 & 12511 & 17351 & 18902 & 42.87 & 21967 & 18911 & 29.45 & 16583 & 17550 & 41.45  & 4.43\% & 7.15\% & 24.51\% & 7.20\% \\ 
        urf4\_187.qasm & 11 & 224028 & 264330 & 378188 & 398902 & 1000.23 & 498085 & 415978 & 2676.34 & 358057 & 363609 & 862.51  & 5.32\% & 8.85\% & 28.11\% & 12.59\% \\ 
        wim\_266.qasm & 11 & 427 & 514 & 717 & 800 & 1.63 & 838 & 766 & 0.76 & 665 & 720 & 1.65  & 7.25\% & 10.00\% & 20.64\% & 6.01\% \\ 
        z4\_268.qasm & 11 & 1343 & 1644 & 2457 & 2648 & 5.72 & 3154 & 2684 & 2.71 & 2308 & 2371 & 5.14  & 6.06\% & 10.46\% & 26.82\% & 11.66\% \\ 
        cycle10\_2\_110.qasm & 12 & 2648 & 3386 & 4791 & 5285 & 12.28 & 5854 & 5105 & 6.56 & 4703 & 4806 & 11.83  & 1.84\% & 9.06\% & 19.66\% & 5.86\% \\ 
        sqrt8\_260.qasm & 12 & 1314 & 1659 & 2335 & 2495 & 11.58 & 3030 & 2619 & 2.39 & 2227 & 2257 & 5.05  & 4.63\% & 9.54\% & 26.50\% & 13.82\% \\ 
        sym9\_146.qasm & 12 & 148 & 127 & 262 & 219 & 0.63 & 317 & 250 & 0.3 & 258 & 193 & 0.69  & 1.53\% & 11.87\% & 18.61\% & 22.80\% \\ 
        plus63mod4096\_163.qasm & 13 & 56329 & 72246 & 101544 & 110144 & 268.3 & 130364 & 111056 & 143.35 & 102491 & 103258 & 252.45  & -0.93\% & 6.25\% & 21.38\% & 7.02\% \\ 
        radd\_250.qasm & 13 & 1405 & 1781 & 2707 & 2955 & 6.13 & 3143 & 2686 & 2.75 & 2491 & 2553 & 5.27  & 7.98\% & 13.60\% & 20.74\% & 4.95\% \\ 
        rd53\_311.qasm & 13 & 124 & 124 & 256 & 283 & 0.62 & 260 & 212 & 0.24 & 246 & 199 & 0.69  & 3.91\% & 29.68\% & 5.38\% & 6.13\% \\ 
        0410184\_169.qasm & 14 & 104 & 104 & 148 & 144 & 0.38 & 170 & 131 & 0.28 & 117 & 110 & 0.23  & 20.95\% & 23.61\% & 31.18\% & 16.03\% \\ 
        cm42a\_207.qasm & 14 & 771 & 940 & 1323 & 1421 & 3.04 & 1683 & 1460 & 2.11 & 1105 & 1202 & 2.49  & 16.48\% & 15.41\% & 34.34\% & 17.67\% \\ 
        plus63mod8192\_164.qasm & 14 & 81865 & 105142 & 145446 & 158513 & 355.55 & 191611 & 161729 & 215.32 & 152014 & 150637 & 380.80  & -4.52\% & 4.97\% & 20.67\% & 6.86\% \\ 
        pm1\_249.qasm & 14 & 771 & 940 & 1348 & 1522 & 3.16 & 1540 & 1408 & 1.46 & 1124 & 1244 & 2.52  & 16.62\% & 18.27\% & 27.01\% & 11.65\% \\ 
        sao2\_257.qasm & 14 & 16864 & 19563 & 28867 & 30118 & 72.63 & 38840 & 30993 & 45.28 & 30736 & 29072 & 78.23  & -6.47\% & 3.47\% & 20.87\% & 6.20\% \\ 
        ham15\_107.qasm & 15 & 3858 & 4819 & 6636 & 7338 & 17.17 & 8608 & 7463 & 7.43 & 6469 & 6834 & 16.68  & 2.52\% & 6.87\% & 24.85\% & 8.43\% \\ 
        misex1\_241.qasm & 15 & 2100 & 2676 & 3343 & 3708 & 11.24 & 4246 & 3910 & 4.06 & 3306 & 3534 & 7.76  & 1.11\% & 4.69\% & 22.14\% & 9.62\% \\ 
        cnt3-5\_180.qasm & 16 & 215 & 209 & 363 & 367 & 0.98 & 456 & 371 & 0.42 & 349 & 313 & 0.98  & 3.86\% & 14.71\% & 23.46\% & 15.63\% \\ 
        inc\_237.qasm & 16 & 4636 & 5863 & 7869 & 8606 & 24.78 & 9420 & 8364 & 8.47 & 7784 & 8272 & 19.90  & 1.08\% & 3.88\% & 17.37\% & 1.10\% \\ 
        mlp4\_245.qasm & 16 & 8232 & 10328 & 15421 & 16547 & 37.96 & 18866 & 15710 & 16.53 & 14870 & 14744 & 35.92  & 3.57\% & 10.90\% & 21.18\% & 6.15\% \\ 
        \textbf{ Geometric mean} & ~ & ~ & ~ & ~ & ~ & ~ & ~ & ~ & ~ & ~ & ~ & ~ & \textbf{ 5.95\%} & \textbf{ 11.88\%} & \textbf{ 23.35\%} & \textbf{ 9.74\%} \\ \hline
    \end{tabular}}
    \end{table*}
    
\begin{figure}[t]
  \centering
  \includegraphics[width=\columnwidth]{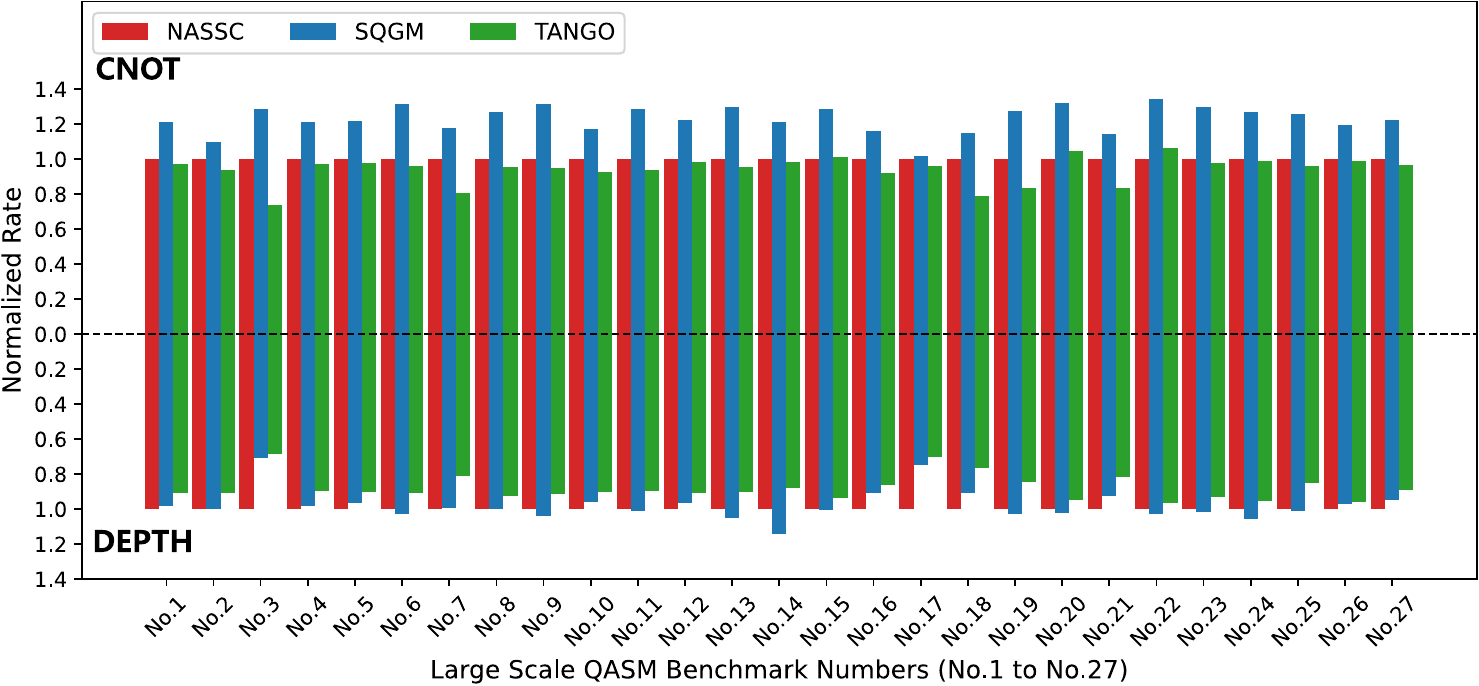}
  \caption{Comparison of total CNOT gates and circuit depth for large-scale quantum circuits on IBM  Q20 (normalize to NASSC)}
  \label{large}
\end{figure}

The experimental results for small-scale quantum circuits are presented in Table. \ref{SMALL} and Fig. \ref{small}. The table provides detailed information on each quantum circuit, including the number of qubits, CNOT gate count, and circuit depth. It also reports the two primary objectives, CNOT gate count and circuit depth, along with the algorithm runtime (in seconds) after applying different mapping algorithms. Additionally, the table highlights the optimization rates of the proposed algorithm in terms of CNOT gate reduction and circuit depth improvement compared to the other two algorithms. Specifically, TANGO achieves a significant reduction in both objectives compared to the NASSC algorithm. For instance, in the circuit $rd32$-$v1\_68$, the gate count is reduced by up to 60\%, and on average, it reduces the gate count by 27.85\% across all small circuits. In terms of circuit depth, Our methodology reduces depth by 50\% in the 
best case, with an average reduction of 25.35\%, indicating notable progress in optimization.

Furthermore, it is worth noting that in most cases, the runtime of the TANGO algorithm is shorter than that of the NASSC algorithm. This is primarily due to the fact that the NASSC algorithm requires the re-synthesis of gate blocks, which incurs high computation. In comparison with the SQGM algorithm, the proposed algorithm also shows strong optimization in both circuit depth and overall gate count. Since the SQGM algorithm primarily focuses on optimizing circuit depth, we observe that its gate count optimization lags behind that of NASSC, despite its superior performance in reducing circuit depth. Specifically, in terms of gate count optimization, our algorithm reduces the gate count by up to 53.45\%, particularly in the circuit $decod24$-$v3\_45$, and on average, it reduces the gate count by 30.92\% across the entire small dataset. For circuit depth optimization, TANGO achieves a reduction of 46.34\% in the best case, particularly in the circuit $3\_17\_13$, and an average reduction of 22.59\% across all small circuits. 

However, in terms of algorithm runtime, the SQGM algorithm outperforms our TANGO algorithm. This is mainly due to the staged selection strategy in the presented algorithm's routing phase, which selects the most suitable SWAP gates, thereby increasing the computational overhead. Therefore, although the runtime of the algorithm is longer than that of SQGM on several quantum circuits, it remains within an acceptable range.

Similarly, good optimization results were also observed in the medium-scale quantum circuits, as shown in the Table. \ref{MIDDLE} and Fig. \ref{middle}, where a certain level of optimization was consistently achieved. Among these circuits, the most significant optimization was seen in the $4gt4$-$v0\_79$ quantum circuit, where TANGO outperformed both the NASSC and SQGM algorithms. In this circuit, it achieved an optimization rate exceeding 30\% in both gate count and circuit depth. However, it is noteworthy that the overall optimization performance of the devised algorithm on medium-scale quantum circuits is lower than that on small quantum circuits. Specifically, in terms of gate count and circuit depth optimization, TANGO improved by 12.92\% and 15.38\%, respectively, compared to the NASSC algorithm. When compared to the SQGM algorithm, our algorithm achieved improvements of 25.69\% and 12.76\% in gate count and circuit depth, respectively.
\begin{table*}[t]
    \centering
    \caption{Comparison of CNOT count and circuit depth between NASSC and SQGM with TANGO on the IBM Rochester for chemistry-related circuits}
    \label{Rochester}
    \resizebox{\linewidth}{!}{
    \begin{tabular}{ccccccccccccccccc}
    \hline
        \multirow{2}*{Benchmark name}  & \multirow{2}*{$\#qubits$} & \multirow{2}*{$CNOT_{total}$} & \multirow{2}*{$Depth_{total}$} & ~ & NASSC & ~ & ~ & SQGM & ~ & ~ & TANGO & ~ & ~ & \multicolumn{2}{c}{Comparison} & ~ \\ 
        ~ & ~ & ~ & ~ & $g_1$ & $d_1$ & $RT$ & $g_2$ & $d_2$ & $RT$ & $g_3$ & $d_3$ & $RT$ & $\Delta_{g_1}$ & $\Delta_{d_1}$ & $\Delta_{g_2}$ & $\Delta_{d_2}$ \\ \hline
        H2\_cmplt\_JW\_ccpvdz.qasm & 20 & 14616 & 16435 & 15934 & 16947 & 33.4 & 19267 & 18908 & 19.68 & 14467 & 16189 & 33.64  & 9.21\% & 4.47\% & 24.91\% & 14.38\% \\ 
        NH\_frz\_JW\_631g.qasm & 20 & 88536 & 95290 & 125540 & 126957 & 372.13 & 125955 & 117404 & 123.73 & 94968 & 101074 & 346.16  & 24.35\% & 20.39\% & 24.60\% & 13.91\% \\ 
        LiH\_cmplt\_JW\_631g.qasm & 22 & 69144 & 74422 & 77302 & 79449 & 238.94 & 104882 & 88937 & 95.09 & 77788 & 80403 & 299.40  & -0.63\% & -1.20\% & 25.83\% & 9.60\% \\ 
        NH\_cmplt\_JW\_631g.qasm & 22 & 173264 & 183892 & 271456 & 271242 & 723.84 & 237428 & 214887 & 265.66 & 211883 & 211028 & 563.42  & 21.95\% & 22.20\% & 10.76\% & 1.80\% \\ 
        C2H4\_cmplt\_JW\_sto3g.qasm & 24 & 640768 & 662500 & 1061760 & 1055078 & 10596.25 & 880689 & 758200 & 1940.35 & 862790 & 805987 & 3094.30  & 18.74\% & 23.61\% & 2.03\% & -6.30\% \\ 
        H2O\_cmplt\_JW\_631g.qasm & 24 & 414240 & 431664 & 627136 & 631831 & 2632.65 & 623973 & 522870 & 2647.83 & 513701 & 524573 & 1625.46  & 18.09\% & 16.98\% & 17.67\% & -0.33\% \\ 
        H2O\_frz\_JW\_631g.qasm & 26 & 244992 & 257780 & 292766 & 285822 & 3470.64 & 366019 & 308337 & 2073.57 & 314626 & 315920 & 928.73  & -7.47\% & -10.53\% & 14.04\% & -2.46\% \\ 
        C2H4\_frz\_JW\_sto3g.qasm & 28 & 312096 & 327532 & 563229 & 548683 & 2051.65 & 444040 & 381175 & 1471.54 & 352665 & 369828 & 1263.66  & 37.39\% & 32.60\% & 20.58\% & 2.98\% \\ 
        LiH\_frz\_JW\_ccpvdz.qasm & 36 & 89080 & 95507 & 194069 & 193187 & 392.65 & 135253 & 113297 & 646.61 & 99368 & 105354 & 459.19  & 48.80\% & 45.47\% & 26.53\% & 7.01\% \\ 
        LiH\_frz\_P\_ccpvdz.qasm & 36 & 86598 & 93681 & 138260 & 135697 & 364.83 & 251250 & 137881 & 117.05 & 149390 & 125184 & 612.24  & -8.05\% & 7.75\% & 40.54\% & 9.21\% \\ 
        LiH\_cmplt\_JW\_ccpvdz.qasm & 38 & 407320 & 419830 & 607302 & 580325 & 2236.02 & 815605 & 554473 & 210.09 & 458819 & 452027 & 1691.23  & 24.45\% & 22.11\% & 43.74\% & 18.48\% \\ 
        H4\_cmplt\_JW\_ccpvdz.qasm & 40 & 479136 & 492286 & 738062 & 721920 & 9527.9 & 829421 & 621305 & 3582.03 & 536780 & 545981 & 1881.04  & 27.27\% & 24.37\% & 35.28\% & 12.12\% \\ 
        H4\_cmplt\_P\_ccpvdz.qasm & 40 & 472132 & 510907 & 777749 & 757353 & 3081.81 & 1121049 & 694728 & 1055.7 & 662832 & 618017 & 3389.60  & 14.78\% & 18.40\% & 40.87\% & 11.04\% \\ 
        \textbf{ Geometric mean} & ~ & ~ & ~ & ~ & ~ & ~ & ~ & ~ & ~ & ~ & ~ & ~ & \textbf{ 17.61\%} & \textbf{17.43\%} & \textbf{25.19\%} & \textbf{7.03\%} \\ \hline
    \end{tabular}    
}
\end{table*}

For large-scale quantum circuits in Table. \ref{LARGE} and Fig. \ref{large}, we used the ultra-large-scale circuit $urf4\_187$, which contains 224,028 two-qubit gates and has a circuit depth of 264,330, as an example. Compared to the NASSC algorithm, the TANGO algorithm achieved a 5.32\% improvement in gate count and an 8.85\% improvement in circuit depth. As previously mentioned, the SQGM algorithm primarily focuses on optimizing circuit depth, resulting in weaker performance in gate count optimization. In contrast, our approach demonstrated a significant improvement of 28.11\% in gate count compared to the SQGM algorithm. However, it is also observed that the optimization rate of the TANGO algorithm on large-scale quantum circuits is lower than on smaller ones, which may indicate that the heuristic algorithm is more likely to get trapped in local optima when searching within a larger solution space. Overall, compared to the NASSC algorithm, the presented algorithm achieved an average optimization rate of 5.95\% in gate count and an 11.88\% improvement in circuit depth. When compared to the SQGM algorithm, it performed excellently in gate count optimization, with a 23.35\% improvement, but showed a slight decrease in circuit depth optimization, achieving a 9.74\% improvement.

\begin{figure}[t]
  \centering
  \includegraphics[width=\columnwidth]{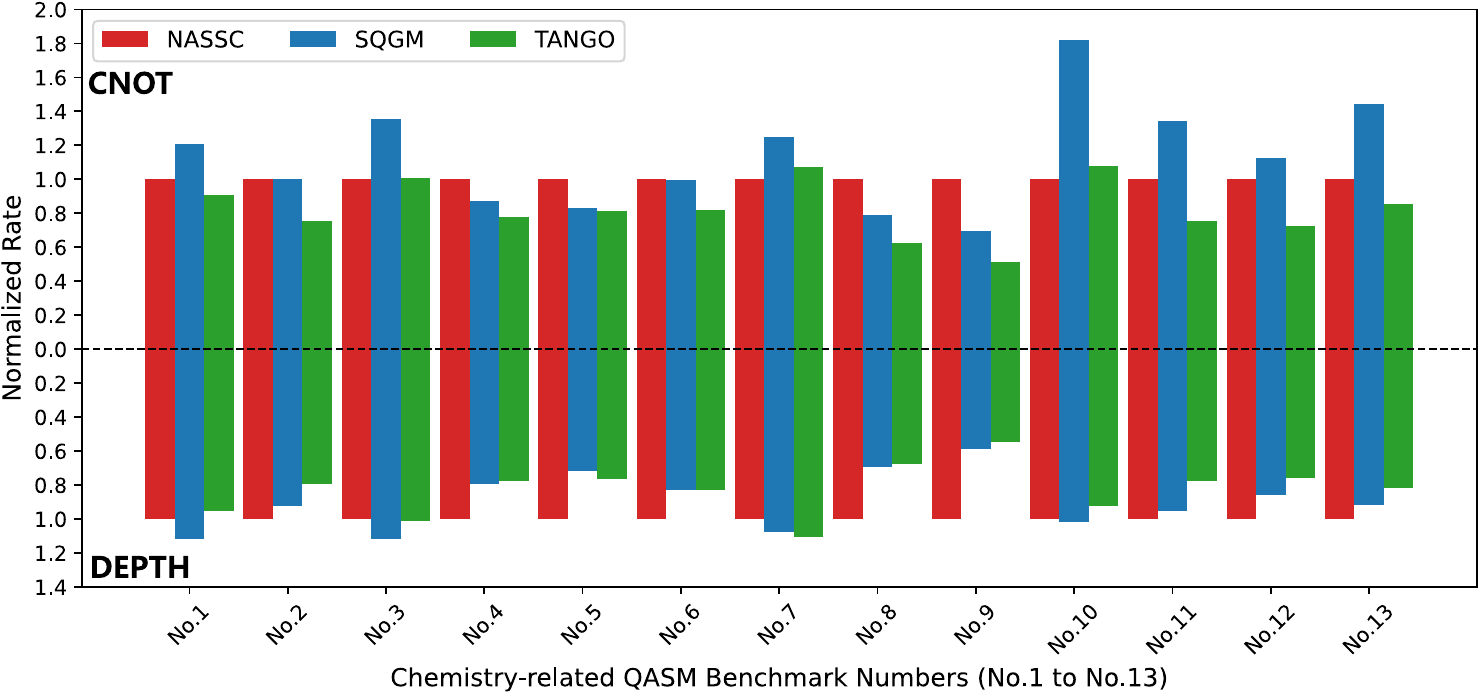}
  \caption{Comparison of total CNOT gates and circuit depth for chemistry-related quantum circuits on IBM Rochester Q53 (normalize to NASSC)}
  \label{rochester}
\end{figure}

\subsection{Comparison with NASSC and SQGM on the Rochester Architecture}
To validate the performance of the algorithm in high-qubit quantum circuits, a set of quantum chemistry-related circuits was selected. These circuits are characterized by a large number of qubits and a high density of quantum gates, making them ideal for testing the algorithm's performance in complex scenarios. For the experiments, the Rochester architecture was chosen for the qubit coupling structure in the experiments. This architecture features relatively sparse connectivity, which typically supports the expansion to a larger number of qubits. As connectivity increases, the interference between qubits also intensifies, making this sparse-connectivity architecture a focal point for researchers.

The experimental results presented in Table. \ref{Rochester} and Fig. \ref{rochester} show that the TANGO algorithm outperforms the NASSC algorithm in 10 out of 13 quantum circuits, demonstrating effective optimization of both gate count and circuit depth. The maximum gate count optimization rate achieved was 48.80\%, observed in the $LiH\_frz\_JW\_ccpvdz$ circuit, with an average optimization rate of 17.61\% across this dataset. Regarding depth optimization, the proposed approach achieved an average optimization rate of 17.43\%. In terms of overall runtime, it outperformed the NASSC algorithm with shorter runtime. However, in the $LiH\_cmplt\_JW\_631g$ and $H2O\_frz\_JW\_631g$ circuits, the devised algorithm showed negative optimization in both gate count and circuit depth, which may be due to the structure of these circuits being more suited to the block re-synthesis module of the NASSC algorithm.

When compared to the SQGM algorithm, the TANGO algorithm consistently outperforms SQGM in gate count optimization across all quantum circuits, achieving an average optimization rate of 25.19\%. However, in terms of circuit depth optimization, our designed algorithm’s performance is less impressive, with a maximum optimization rate of 18.48\% and an average rate of 7.03\%. This result indirectly suggests that reducing the number of gates in a circuit does not always lead to a significant reduction in its depth. Therefore, finding a balance between gate count and circuit depth, such that both optimization objectives fall within an acceptable range, is the key challenge addressed in this paper.

\section{conclusion}\label{sec5}
In this paper, we propose a quantum circuit mapping algorithm called TANGO, which jointly optimizes both circuit depth and gate count. Specifically, during the initial mapping phase, we design a dual-factor initial mapping algorithm that considers the influence between mapped and unmapped qubits and defines relevant parameters to balance these two components, thus improving the performance of initial mapping. In the qubit routing phase, we propose a two-stage qubit routing algorithm. Using the number of executable gates as the primary evaluation metric, and combining quantum gate distance, circuit depth, and bidirectional-look SWAP strategy, the method effectively identifies the SWAP gates with optimal exchange efficiency and minimal circuit depth. Finally, by integrating advanced quantum gate optimization techniques, the total gate count and depth of the circuit are further reduced. Compared to existing state-of-the-art algorithms, our algorithm achieves significant optimizations in both gate count and circuit depth across different datasets and architectures, with no substantial increase in runtime. These results demonstrate the scalability and adaptability of our algorithm, offering a viable solution to the quantum mapping problem.

\section*{acknowledgments}
This research was supported by the National Nature Science Foundation of China (Grant No. 62101600, Grant No. 62471070), and the State Key Lab of Processors, Institute of Computing Technology, CAS (Grant No. CLQ202404).
% \section*{Acknowledgment}

% \begin{thebibliography}{00}

% \bibitem{b7} M. Young, The Technical Writer's Handbook. Mill Valley, CA: University Science, 1989.
% \end{thebibliography}
\vspace{12pt}
% IEEE conference templates contain guidance text for composing and formatting conference papers. Please ensure that all template text is removed from your conference paper prior to submission to the conference. Failure to remove the template text from your paper may result in your paper not being published.

%\clearpage
\bibliographystyle{IEEEtran}
%\bibliography{IEEEabrv,ref}

\end{document}